\documentclass[]{article}

\usepackage{amssymb}
\usepackage{amsmath}
\usepackage{caption}
\usepackage{subcaption}
\usepackage{algorithm2e}
\usepackage{url}
\usepackage{cleveref}
\usepackage{tabularx}
\usepackage[smaller,nolist]{acronym}
\usepackage{authblk}
\usepackage{graphicx}

\usepackage[left=3.5cm, right=3.5cm]{geometry}

\begin{acronym}
    \acro{AI}{Artificial Intelligence}
    \acro{RS}{Recommender System}
    \acro{CF}{Collaborative Filtering}
    \acro{DL}{Deep Learning}
    \acro{MF}{Matrix Factorization}
    \acro{CF4J}{Collaborative Filtering for Java}
    \acro{NN}{Neural Network}
    \acro{KNN}{k-Nearest Neighbors}
    \acro{BeMF}{Bernoulli Matrix Factorization}
    \acro{DirMF}{Dirichlet Matrix Factorization}
    \acro{ResBeMF}{Restricted Bernoulli Matrix Factorization}
    \acro{MAE}{Mean Absolute Error}
\end{acronym}

\title{Incorporating Recklessness to Collaborative Filtering based Recommender Systems}

\author[1,2]{Diego Pérez-López}
\author[1,3]{Fernando Ortega}
\author[1,4,5]{Ángel González-Prieto}
\author[1,2]{Jorge Dueñas-Lerín}

\affil[1]{KNODIS Research Group, Universidad Politécnica de Madrid}
\affil[2]{Universidad Politécnica de Madrid}
\affil[3]{Departamento de Sistemas Informáticos, Universidad Politécnica de Madrid}
\affil[4]{Departamento de Álgebra, Geometría y Topología, Universidad Complutense de Madrid}      
\affil[5]{Instituto de Ciencias Matemáticas (CSIC-UAM-UCM-UC3M)}  

\date{}

\begin{document}

\maketitle

\begin{abstract}
Recommender systems are intrinsically tied to a reliability/coverage dilemma: The more reliable we desire the forecasts, the more conservative the decision will be and thus, the fewer items will be recommended. This causes a detriment to the predictive capability of the system, as it is only able to estimate potential interest in items for which there is a consensus in their evaluation, rather than being able to estimate potential interest in any item.
In this paper, we propose the inclusion of a new term in the learning process of matrix factorization-based recommender systems, called recklessness, that takes into account the variance of the output probability distribution of the predicted ratings. In this way, gauging this recklessness measure we can force more spiky output distribution, enabling the control of the risk level desired when making decisions about the reliability of a prediction. Experimental results demonstrate that recklessness not only allows for risk regulation but also improves the quantity and quality of predictions provided by the recommender system.
\end{abstract}


\section{Introduction} \label{sec:intro}

\Acp{RS} have become a powerful tool for an internet-based society in recent years. Their main purpose is to provide users with a series of products or services that may be of interest to them based on their past preferences. Therefore, \acp{RS} have applications in fields such as e-commerce~\cite{ge2020understanding,alamdari2020systematic}, online entertainment platforms~\cite{kumar2020movie,cheuque2019recommender,fessahaye2019t}, or smart home automation~\cite{quijano2020recommender,jannach2021survey}, to name a few examples.

Due to their popularity, research on \acp{RS} has grown significantly in the last decade. Most of these studies have focused on improving the quantitative aspects of the predictions generated by the artificial intelligence models on which \acp{RS} are based~\cite{wu2022survey,koren2021advances}. However, there are also other studies that aim to improve the qualitative aspects of these systems~\cite{bertani2020combining,isufi2021accuracy,pujahari2023ordinal}.

A recent study~\cite{ortega2021providing} provides a significant change in the operation of \ac{CF} based \acp{RS}, the most popular type of \ac{RS}~\cite{bobadilla2013recommender}. Traditionally, \ac{CF} has approached the calculation of its predictions as a regression problem. However, most datasets used to train \ac{CF} have a discrete rating system (for example, from 1 to 5 stars), and therefore, predictions should be modelled as a classification problem.

This change means that the predictions of a \ac{CF} based \acp{RS} are no longer real numbers but discrete probability distributions that represent how likely a user is to rate an item with a score. For example, instead of predicting that item $i$ will interest user $u$ with 4.33 stars, a classification-based recommendation system returns the probability \texttt{<0, 0, 0, 0.66, 0.34>}, assuming that the possible score range is from 1 to 5 stars. For this reason, these \acp{RS} are known as probability-based \ac{RS}.

Thanks to this new form of prediction, it is possible to know the reliability of a prediction: Following the previous example, we will know that a prediction of \texttt{<0, 0, 0, 0, 1>} is much more reliable than one of \texttt{<0.19, 0.2, 0.2, 0.2, 0.21>}, even though they have the same mode. By using this reliability, it is possible to gauge the output of a \ac{RS} to offer more reliable predictions at the expense of sacrificing less reliable ones. In other words, the quantity and quality of predictions can be balanced by setting a threshold from which a prediction is considered not reliable enough. However, this mechanism also leads to an undesired effect: the system tends to output almost-flat predictions, with small deviations from the uniform distribution, and only a few clear items are distinguished with a more spiky distribution. This produces a more conservative system that, even with very small thresholds, suffers a noticeable drop in the coverage of items to be recommended.

To address this problem, in this paper we propose the inclusion of a new regularization term in the cost function of \ac{CF} models driven by probability-based \ac{RS}, called recklessness, which aims to force the model to find solutions with higher quality predictions or a greater number of predictions. This is a hyper-parameter of the system that impacts the training process of the \ac{RS} fostering non-uniform distributions with higher peaks. Empirical results show that, gauging properly this hyper-parameter, we can push the Pareto front to get better predictors, both in terms of coverage and accuracy.

The rest of the article is structured as follows: \Cref{sec:related-work} dives into the related work, \Cref{sec:recklessness} formalizes the new recklessness regularization proposed in this manuscript, \Cref{sec:experiments} shows the experimental results of the proposed regularization, and \cref{sec:conclusions} presents the conclusions and future work of this research.

\section{Related Work} \label{sec:related-work}

Probability-based \ac{RS} have sparked the interest of the research community in recent years. As mentioned in the previous section, the most representative work in this area is~\cite{ortega2021providing}, where the authors propose encoding user ratings using one-hot encoding and then factorizing the \ac{CF} voting matrix based on the Bernoulli distribution. In the same vein, the authors suggest DirMF~\cite{lara2022dirichlet}, another matrix factorization model but based on the Dirichlet distribution instead of Bernoulli. OrdRec~\cite{koren2011ordrec} proposes a model that produces user-item scores converted into probabilities over an ordinal set of ratings. URP~\cite{marlin2003modeling} introduces another probability-based recommendation system encoded as a generative latent variable model for rating-based collaborative filtering. In a similar spirit, \cite{pujahari2022handling} further explores this idea and, instead of using point estimation tools to decide the parameters of the model, such as the likelihood of the predictions, the authors propose to apply distribution estimation. Finally, \cite{bobadilla2022neural} presents a collaborative filtering approach based on artificial neural networks, adding a softmax output layer to provide probabilities instead of numerical predictions.

The increased interest in probability-based \ac{RS} lies in the reliability associated with each computed prediction. The reliability of \ac{RS} has been extensively studied. On one hand, some authors aim to add value to the reliability of individual predictions. For example, \cite{ahmadian2019novel,ahmadian2022alleviating} employ reliability to address the issue of voting matrix sparsity in \ac{CF}, while \cite{alonso2019robust} suggests using reliability to detect shilling attacks in \ac{RS}. On the other hand, the overall reliability of \ac{RS} is also investigated. In this regard, \cite{marshall2016mood} explores the mood sensitivity aspect of truth discovery to build a reliable \ac{RS}, \cite{hou2018reliable} develops a reliable and private \ac{RS} for applying \ac{CF} in medical environments, and \cite{yuan2018ms} considers utilizing user reputation to resist malicious users and ensure the recommendation's reliability.

As mentioned earlier, this article aims to provide \ac{RS} with a mechanism that allows adjusting the reliability of their predictions. To achieve this, a new hyper-parameter is introduced into the model's learning process. A key point of this work is that the new hyper-parameter is understandable for humans. As we will see later on, positive values provide \ac{RS} with less reliable predictions, whereas negative values provide more reliable predictions. This fact allows humanizing the \ac{RS} and assisting their end-users. Humanizing \ac{AI}~\cite{israni2019humanizing,robert2017growing} is one of the major challenges faced by science and would help address its application in medical~\cite{hamet2017artificial}, ethical~\cite{russell2015ethics} and customer service~\cite{schanke2021estimating} issues, among others.

\section{Recklessness Regularization} \label{sec:recklessness}

The general setup of a probability-based \ac{RS} is the following. Suppose that our system contains $U$ users and $I$ items, so that we can collect their ratings in a (highly) sparse matrix $R = (r_{ui})$ of order $U \times I$, where $r_{ui}$ is the rating that the user $u$ assigned to the item $i$. Here, $r_{ui}$ belongs to some finite set $\Omega$ of possible ratings (typically, $\Omega = \{1, \ldots, 5\}$ for $5$-point scales or $\Omega = \{0,1\}$ for like/dislike ratings). If the rating of the user $u$ to the item $i$ is missing, we will denote it by $r_{ui} = \bullet$.

In this setting, a probability-based \ac{RS} is a function that, to any user-item pair $(u,i)$, assigns a probability distribution $\hat{\mathbb{P}}_{ui}$ on $\Omega$, seen as a sample space. In this sense, $\hat{\mathbb{P}}_{ui}(s) \in [0,1]$ should be thought as the estimation of the probability that the user $u$ assigns to item $i$ a rating $s$. From this probability, the prediction $\hat{r}_{ui} \in \Omega$ can be computed as the mode of this distribution, and its reliability $\hat{\rho}_{ui} \in [0,1]$ as the probability of this mode. In this fashion, if we set a reliability threshold $0 \leq \theta \leq 1$, no prediction is issued if $\hat{\rho}_{ui} \leq \theta$.

However, more sophisticated statistics of the estimated probability $\hat{\mathbb{P}}_{ui}$ can be calculated. In particular, we can consider a random variable $X: \Omega \to \mathbb{R}$ and compute the variance $\mathbb{V}^X_{\hat{\mathbb{P}}_{ui}} = \mathbb{E}(X^2) - \mathbb{E}(X)^2$ of this random variable.
For applications, typically the set $\Omega$ of possible scores is a set of finitely many numbers $\Omega = \{x_1, \ldots, x_s\}$ (for instance $\Omega = \{1,2,3,4,5\}$ in MovieLens~\cite{harper2015movielens}), and we can take $X : \Omega \to \mathbb{R}$ to be the identity map $X(x_i)=x_i$. In this case, its variance is given by
\begin{equation}
    \mathbb{V}^X_{\hat{\mathbb{P}}_{ui}} = \mathbb{E}(X^2) - \mathbb{E}(X)^2 = \sum_{k = 1}^s x_k^2 \hat{\mathbb{P}}_{ui}(x_k) - \left(\sum_{k = 1}^s x_k \hat{\mathbb{P}}_{ui}(x_k)\right)^2.
\end{equation}

Now, let us suppose that the estimated probability $\hat{\mathbb{P}}_{ui} = \hat{\mathbb{P}}_{ui}(p_u,q_i)$ depends on some parameters $p_u, q_i \in \mathbb{R}^D$ associated to each user $u$ and item $i$. In the field of \acp{RS}, these parameters are typically referred to as the hidden factors. Let us collect these parameters by columns into matrices $P = (p_1\,|\, \ldots\,|\, p_U)$ and $Q = (q_1\,|\, \ldots \,|\, q_I)$ of sizes $D \times U$ and $D \times I$, respectively.

To train these parameters, a cost function $\mathcal{F}: \mathbb{R}^{D \times U} \times \mathbb{R}^{D \times I} \to \mathbb{R}$ is sought to be optimized so that the optimal parameters are
\begin{equation}
    (P^*, Q^*) = \underset{P, Q}{\textrm{argmin}}\,\mathcal{F}(P, Q).
\end{equation}
This optimal point is searched using some classical optimization algorithm such as stochastic gradient descent. Once computed this minimum, if $P^* = (p_u^*)$ and $Q^* = (q_i^*)$, the estimated probability distribution for a pair $(u,i)$ is $\hat{\mathbb{P}}_{u,i}(p_u^*, q_i^*)$.

Our proposal in this work is to modify this cost function to take into account the variance of the predicted distribution obtained. In this way, fixed a real value $\alpha \geq 0$, we define the modified cost function
\begin{equation}
    \mathcal{F}_\alpha(P,Q) = \mathcal{F}(P,Q) - \alpha \sum_{u,i} \hat{\mathbb{V}}^X_{\hat{\mathbb{P}}_{u,i}(p_u^*, q_i^*)}.
\end{equation}
In the case that the random variable $X: \Omega = \{x_1, \ldots, x_s\} \to \mathbb{R}$ is the identity map, then
\begin{equation}
\mathcal{F}_\alpha(P,Q) = \mathcal{F}(P,Q) - \alpha \sum_{u,i}\left(\sum_{k = 1}^s x_k^2 \hat{\mathbb{P}}_{ui}(x_k) - \left(\sum_{k = 1}^s x_k \hat{\mathbb{P}}_{ui}(x_k)\right)^2\right)
\end{equation}

Observe that, the larger the value of $\alpha$, the more important will be to maximize the variance of the output. In this manner, large values of $\alpha$ will tend to provide estimated probability distributions $\hat{\mathbb{P}}_{ui}$ with a large variance. In other words, the algorithm is forced to return a spiky output distribution, with sharp maxima and minima. In this sense, we are impelling the method to take a risk and to stress the mode of the distribution. For this reason, we shall refer to $\alpha$ as the \textit{recklessness} parameter.

To minimize it, we shall follow a stochastic gradient descend on items and users. A straightforward calculation shows that the derivatives are given by
\begin{align}
    \begin{split}
    \partial_{p_{u}}\mathcal{F}_\alpha = &\; \partial_{p_u} \mathcal{F} - \alpha \sum_{i,k} x_k^2 \partial_{p_{u}}\hat{\mathbb{P}}_{ui}(x_k) \\
    & + 2\alpha \sum_i \left(\sum_{k} x_k \hat{\mathbb{P}}_{ui}(x_k)\right)\left(\sum_{k} x_k \partial_{p_{u}}\hat{\mathbb{P}}_{ui}(x_k)\right),\\
    \end{split}
\end{align}
\begin{align}
    \begin{split}
    \partial_{q_{i}}\mathcal{F}_\alpha = &\; \partial_{q_i} \mathcal{F} - \alpha \sum_{u,k} x_k^2 \partial_{q_{i}}\hat{\mathbb{P}}_{ui}(x_k) \\
    & + 2\alpha \sum_u\left(\sum_{k} x_k \hat{\mathbb{P}}_{ui}(x_k)\right)\left(\sum_{k} x_k \partial_{q_{i}}\hat{\mathbb{P}}_{ui}(x_k)\right).
    \end{split}
\end{align}
Here $\partial_{p_u} \mathcal{F}$ and $\partial_{q_i} \mathcal{F}$ are the gradients of the standard cost function of the given algorithm.

Therefore, the modified update rules for stochastic gradient descend with learning rate $\eta > 0$ for the pair user-item $(u,i)$ with $r_{ui} \neq \bullet$ are
\begin{align}
    \begin{split}
    p_u \leftarrow &\; p_u - \eta \partial_{p_u} \mathcal{F} + \alpha \eta \sum_{k} x_k^2 \partial_{p_{u}}\hat{\mathbb{P}}_{ui}(x_k) \\
    & - 2\alpha \eta\left(\sum_{k} x_k \hat{\mathbb{P}}_{ui}(x_k)\right)\left(\sum_{k} x_k \partial_{p_{u}}\hat{\mathbb{P}}_{ui}(x_k)\right),\\
    \end{split}
\end{align}
\begin{align}
    \begin{split}
    q_i \leftarrow &\; q_i - \eta\partial_{q_i} \mathcal{F} + \alpha \eta\sum_{k} x_k^2 \partial_{q_{i}}\hat{\mathbb{P}}_{ui}(x_k) \\
    & - 2\alpha \eta\left(\sum_{k} x_k \hat{\mathbb{P}}_{ui}(x_k)\right)\left(\sum_{k} x_k \partial_{q_{i}}\hat{\mathbb{P}}_{ui}(x_k)\right).
    \end{split}
\end{align}

To relate this with the usual \ac{RS} models, the standard procedure is to suppose that $\hat{\mathbb{P}}_{ui}$ has some prescribed distribution, such as a normal distribution \cite{mnih2007probabilistic}, a (multiple) Bernoulli distribution \cite{ortega2021providing}, the Dirichlet distribution \cite{lara2022dirichlet} or a restricted binomial distribution \cite{gonzalez2022resbemf}. In this manner, the cost function $\mathcal{F}$ is given by the minus log-likelihood of the known ratings
\begin{equation}
    \mathcal{F}(P, Q) = -\log \mathcal{L}(P,Q) = - \log \prod_{r_{ui} \neq \bullet} \hat{\mathbb{P}}_{ui}(r_{ui}) = -\sum_{r_{ui} \neq \bullet} \log\hat{\mathbb{P}}_{ui}(r_{ui}).
\end{equation}
In this way, the recklessness cost function would be
\begin{equation}    
    \mathcal{F}_\alpha(P, Q) = -\sum_{r_{ui} \neq \bullet} \log\hat{\mathbb{P}}_{ui}(r_{ui}) - \alpha \sum_{u,i,k} x_k^2 \hat{\mathbb{P}}_{ui}(x_k) + \alpha \left(\sum_{u,i,k} x_k \hat{\mathbb{P}}_{ui}(x_k)\right)^2.
\end{equation}
Here, recall that the dependence of this function on the parameters $P$ and $Q$ is in the predicted probability  $\hat{\mathbb{P}}_{ui} = \hat{\mathbb{P}}_{ui}(p_u,q_i)$. In this way, the usual gradients of the standard cost function is given by
\begin{align}
\partial_{p_u} \mathcal{F} = -\sum_{r_{ui} \neq \bullet} \frac{\partial_{p_{u}}\hat{\mathbb{P}}_{ui}(r_{ui})}{\hat{\mathbb{P}}_{ui}(r_{ui})}, \quad
\partial_{q_i} \mathcal{F} = -\sum_{r_{ui} \neq \bullet} \frac{\partial_{q_{i}}\hat{\mathbb{P}}_{ui}(r_{ui})}{\hat{\mathbb{P}}_{ui}(r_{ui})}.
\end{align}

Furthermore, if we suppose that the parameters of this distribution depend on the latent factors $p_u$ and $q_i$ through the inner product $p_u \cdot q_i$, we have that $\hat{\mathbb{P}}_{u,i}(x) = f_x(p_u \cdot q_i)$ for some function $f$ depending on the sample $x$. 
Therefore, the corresponding updating rules for a pair $(u,i)$ with $r_{ui} \neq \bullet$ are
\begin{align}
    \begin{split}
    p_u \leftarrow &\; p_u + \eta \frac{\partial_{p_{u}}\hat{\mathbb{P}}_{ui}(r_{ui})}{\hat{\mathbb{P}}_{ui}(r_{ui})} + \alpha \eta \sum_{k} x_k^2 f'_{x_k}(p_u \cdot q_i)q_i \\
    & - 2\alpha \eta\left(\sum_{k} x_k f_{x_k}(p_u \cdot q_i)\right)\left(\sum_{k} x_k f'_{x_k}(p_u \cdot q_i)q_i\right),
    \end{split}
\end{align}
\begin{align}
    \begin{split}
    q_i \leftarrow &\; q_i + \eta \frac{\partial_{q_{i}}\hat{\mathbb{P}}_{ui}(r_{ui})}{\hat{\mathbb{P}}_{ui}(r_{ui})} + \alpha \eta \sum_{k} x_k^2 f'_{x_k}(p_u \cdot q_i)p_u \\
    & - 2\alpha \eta\left(\sum_{k} x_k f_{x_k}(p_u \cdot q_i)\right)\left(\sum_{k} x_k f'_{x_k}(p_u \cdot q_i)p_u\right).
    \end{split}
\end{align}
These are the new updating rules for the latent factors using a gradient descend-based training for the \ac{RS} model.

\section{Experimental Evaluation} \label{sec:experiments}

In this section, we present the experimental results that verify the influence of the proposed recklessness regularization on probability-based (\ac{RS}). Specifically, the experiments will be conducted using the \ac{BeMF} \cite{ortega2021providing} recommendation model. This model has been chosen as the probability-based \ac{RS} that provides the most accurate predictions.

As defined in the previous section, to incorporate recklessness regularization into a probability-based \ac{RS}, it is necessary to modify its cost function. In the case of \ac{BeMF}, it is based on parameters $P = (p_u^k)$ for each user $u$ and possible rating $x_k$ and $Q = (q_i^k)$ for each item $i$ and rating $x_k$.
Therefore, as proposed in \cite{ortega2021providing}, the standard cost function without recklessness regularization for \ac{BeMF} is
\begin{equation}
    \mathcal{F}(P,Q) = - \sum_{r_{ui} = x_k} \log(\sigma(p_u^k \cdot q_i^k)) - \sum_{r_{ui} \neq x_k, \bullet} \log(1-\sigma(p_u^k \cdot q_i^k)),
\end{equation}
where $\sigma: \mathbb{R} \to [0,1]$, $\sigma(x) = (1+e^{-x})^{-1}$ is the logistic function. The predicted probability is thus
\begin{equation}
    \hat{\mathbb{P}}_{ui}(x_k) = \textrm{sm}_k(p_u^1 \cdot q_i^1, \ldots, p_u^n \cdot q_i^n) = \frac{\sigma(p_u^k \cdot q_i^k)}{\sum_l \sigma(p_u^l \cdot q_i^l)},
\end{equation}
where $\textrm{sm}_k(s_1, \ldots, s_n) = \sigma(s_k)/\sum_l \sigma(s_l)$ is the $k$-th entry of the softmax function. To simplify notation, we shall denote $\textrm{sm}_k(p_u \cdot q_i) = \textrm{sm}_k(p_u^1 \cdot q_i^1, \ldots, p_u^n \cdot q_i^n)$

In this way, if we modify the cost function with the recklessness parameter $\alpha$ we get
\begin{align}
    \begin{split}
    \mathcal{F}_\alpha(P, Q) = &\,  - \sum_{r_{ui} = x_k} \log(\sigma(p_u^k \cdot q_i^k)) - \sum_{r_{ui} \neq x_k, \bullet} \log(1-\sigma(p_u^k \cdot q_i^k)) \\
    & \qquad - \alpha \sum_{u,i,k} x_k^2 \textrm{sm}_k(p_u \cdot q_i) + \alpha \left(\sum_{u,i,k} x_k\textrm{sm}_k(p_u \cdot q_i)\right)^2.
    \end{split}
\end{align}

As shown in \cite{ortega2021providing}, the corresponding gradients for the standard cost function are given by 
\begin{align}
    \partial_{p_u^k}\mathcal{F}(P,Q) = & -\sum_{r_{u,i} = x_k} {(1-\sigma(p_{u}^k\cdot q_i^k))}q_{i}^k + \sum_{r_{u,i} \neq x_k,\bullet} {\sigma(p_{u}^k\cdot q_i^k)}q_{i}^k,
\end{align}
\begin{align}
    \partial_{q_u^k}\mathcal{F}(P,Q) = &-\sum_{r_{u,i} = x_k} {(1-\sigma(p_{u}^k\cdot q_i^k))}p_{u}^k +  \sum_{r_{u,i} \neq x_k, \bullet} {\sigma(p_{u}^k\cdot q_i^k)}p_{u}^k.
\end{align}
Therefore, the gradients for the recklessness cost function are
\begin{align}
    \begin{split}
    \partial_{p_u^k}\mathcal{F}_\alpha(P,Q) = &- \sum_{r_{u,i} = x_k} {(1-\sigma(p_{u}^k\cdot q_i^k))}q_{i}^k + \sum_{r_{u,i} \neq x_k,\bullet} {\sigma(p_{u}^k\cdot q_i^k)}q_{i}^k\\
    &\quad+ \alpha \eta \sum_{i,t} x_t^2 \partial_{k}\textrm{sm}_{t}(p_u \cdot q_i)q_i^t \\
    &\quad- 2\alpha \eta \sum_i\left(\sum_{t} x_t \textrm{sm}_t(p_u \cdot q_i)\right)\left(\sum_{t} x_t \partial_{k}\textrm{sm}_{t}(p_u \cdot q_i)q_i^t\right), \\
    \end{split}
\end{align}
\begin{align}
    \begin{split}
    \partial_{q_i^k}\mathcal{F}_\alpha(P,Q) = &-\sum_{r_{u,i} = x_k} {(1-\sigma(p_{u}^k\cdot q_i^k))}p_{u}^k +  \sum_{r_{u,i} \neq x_k, \bullet} {\sigma(p_{u}^k\cdot q_i^k)}p_{u}^k \\
    &\quad+ \alpha \eta \sum_{u,t} x_t^2 \partial_{k}\textrm{sm}_{t}(p_u \cdot q_i)p_u^t \\
    &\quad- 2\alpha \eta\sum_u\left(\sum_{t} x_t \textrm{sm}_t(p_u \cdot q_i)\right)\left(\sum_{t} x_t \partial_{k}\textrm{sm}_{t}(p_u \cdot q_i)p_u^t\right).
    \end{split}
\end{align}
where recall that $\textrm{sm}_{t}(p_u \cdot q_i) = \sigma(p_u^t \cdot q_i^t)/{\sum_l \sigma(p_u^l \cdot q_i^l)}$ is the softmax function and $\partial_k\textrm{sm}_{t}(p_u \cdot q_i) = \textrm{sm}_{t}(p_u \cdot q_i)\left(\delta_{tk} - \textrm{sm}_{k}(p_u \cdot q_i)\right)$, being $\delta_{tk} = 1$ if $t = k$ and $\delta_{tk} = 0$ if $t \neq k$.

Hence, the updating rule for a pair user-item $(u,i)$ with $r_{ui} = x_k$ are 
\begin{align}
    \begin{split}        
    p_u^k \leftarrow & p_u^k + (1-\sigma(p_{u}^k\cdot q_i^k))q_{i}^k - \alpha \eta \sum_{t} x_t^2 \partial_{k}\textrm{sm}_{t}(p_u \cdot q_i)q_i^t \\
    &\quad+ 2\alpha \eta \left(\sum_{t} x_t \textrm{sm}_t(p_u \cdot q_i)\right)\left(\sum_{t} x_t \partial_{k}\textrm{sm}_{t}(p_u \cdot q_i)q_i^t\right),\\
    \end{split}
\end{align}
\begin{align}
    \begin{split}        
    q_i^k \leftarrow & q_i^k + (1-\sigma(p_{u}^k\cdot q_i^k))p_{u}^k - \alpha \eta \sum_{t} x_t^2 \partial_{k}\textrm{sm}_{t}(p_u \cdot q_i)p_u^t \\
    &\quad+ 2\alpha \eta\left(\sum_{t} x_t \textrm{sm}_t(p_u \cdot q_i)\right)\left(\sum_{t} x_t \partial_{k}\textrm{sm}_{t}(p_u \cdot q_i)p_u^t\right).
    \end{split}
\end{align}
In the same spirit, if $r_{u,i} \neq x_k$, then
\begin{align}
    \begin{split}
    p_u^k \leftarrow & p_u^k - \sigma(p_{u}^k \cdot q_i^k)q_{i}^k - \alpha \eta \sum_{t} x_t^2 \partial_{k}\textrm{sm}_{t}(p_u \cdot q_i)q_i^t \\
    &\quad+ 2\alpha \eta \left(\sum_{t} x_t \textrm{sm}_t(p_u \cdot q_i)\right)\left(\sum_{t} x_t \partial_{k}\textrm{sm}_{t}(p_u \cdot q_i)q_i^t\right),\\
    \end{split}
\end{align}
\begin{align}
    \begin{split} 
    q_i^k \leftarrow & q_i^k - \sigma(p_{u}^k\cdot q_i^k)p_{u}^k  - \alpha \eta \sum_{t} x_t^2 \partial_{k}\textrm{sm}_{t}(p_u \cdot q_i)p_u^t \\
    &\quad+ 2\alpha \eta\left(\sum_{t} x_t \textrm{sm}_t(p_u \cdot q_i)\right)\left(\sum_{t} x_t \partial_{k}\textrm{sm}_{t}(p_u \cdot q_i)p_u^t\right).
    \end{split}
\end{align}

Notice that \ac{BeMF} can be complemented with a regularization term to improve the convergence of the method. In that case, an extra term $-\beta p_u^k$ must be added to the updating rule for $p_u^k$ and a term $-\beta q_i^k$ must be added to update $q_i^k$, where $\beta \geq 0$ is the regularization hyper-parameter.

\subsection{Experimental Environment Definition}

Regarding the experimental environment, we have employed the same configuration as the majority of research studies in the field of \ac{RS}.

The implementation of the recommendation models has been carried out using the Java's CF4J framework~\cite{ortega2018cf4j,ortega2021providing}. Additionally, Python's libraries Pandas and MatPlotLib has been used for figures generation. The source code for all the conducted experiments is publicly available on GitHub\footnote{\url{https://github.com/KNODIS-Research-Group/recklessness-regularization}} to ensure result transparency and facilitate the reproducibility of the experiments. This source code has been run on a server with 2 x Intel(R) Xeon(R) Gold 6230 CPU @ 2.10 GHz (40 cores / 80 threads) and 256 GB of RAM.

The datasets used in these experiments to evaluate the recklessness regularization are FilmTrust~\cite{guo2013novel} and MovieLens~\cite{harper2015movielens} (this later dataset, both in its 100K and 1M sizes). The main features of these datasets are shown in \cref{tab:datasets}. Remark that we used the training and test partitions provided by CF4J. Larger datasets, such as MyAnimeList or Netflix Prize, were not employed due to the significant computational cost involved in tuning the model hyper-parameters for these datasets.

\begin{table}[ht]
\centering
\scriptsize
\begin{tabularx}{\textwidth}{|l|X|X|X|X|X|}
    \hline
    Dataset        & Number of users & Number of items & Number of ratings & Number of test ratings & Scores     \\ \hline
    FilmTrust      & 1,508           & 2,071           & 32,675            & 2,819                  & 0.5 to 4.0 \\ \hline
    MovieLens100K  & 943             & 1,682           & 92,026            & 7,974                  & 1 to 5     \\ \hline
    MovieLens1M    & 6,040           & 3,706           & 911,031           & 89,178                 & 1 to 5     \\ \hline
\end{tabularx}
\caption{Main parameters of the datasets used in the experiments sorted by number of ratings.}
\label{tab:datasets}
\end{table}

The evaluation of the recommendation model has been carried out considering the peculiarities offered by probability-based \ac{RS}. As defined previously, this type of \ac{RS} provides not only the predictions of ratings but also their reliability. This means that it is possible to modulate the recommender's output to include more or less reliable predictions. As a consequence, the quantity and quality of these predictions will vary. On the one hand, if the reliability threshold used to filter predictions is high, there will be fewer predictions, but they will be very accurate. On the other hand, if the same threshold is set low, there will be many predictions, but they may be less reliable. This fact transforms the evaluation of probability-based \ac{RS} into a multi-objective optimization problem in which both the quality and quantity of predictions are maximized.

In this regard, we are obligated to define two quality measures to evaluate the performance of the recommender. One to assess the accuracy of the predictions and the other to evaluate the quantity of the predictions. In this vein, given a threshold $\theta$ we consider the set
\begin{equation}
    T^\theta = \{(u,i) \in R^{\textrm{test}} | \hat{\rho}_{u,i} \geq \theta\}
\end{equation}
of pairs user $u$ and item $i$ in the test split $R^{\textrm{test}}$ for which the reliability of the predictions is equal or greater than $\theta$. With it, we define
\begin{equation}
\textrm{MAE}^\theta = \frac{1}{\#T^\theta} \sum_{(u,i) \in T^\theta} \frac{|r_{u,i} - \hat{r}_{u,i}|}{max(\Omega) - min(\Omega)}
\end{equation}
to measure the quality of the predictions as the normalized \ac{MAE} of the predictions with a reliability greater or equal than $\theta$, and
\begin{equation}
\textrm{coverage}^\theta = \frac{\#T^\theta}{\#R^{\textrm{test}}}
\end{equation}
to measure the proportion of the predictions with a reliability greater or equal than $\theta$ with respect to a test split $R^{\textrm{test}}$.

However, these measures report the quality of the model for a fixed reliability threshold $\theta$. To evaluate the real quality of the model, we must average these measures for different values of $\theta$. As $\theta \in [0,1]$, we sample $\theta$ in an equidistant partition of the unit interval with $N$ points
\begin{equation}
\theta_k = \frac{k}{N-1}
\end{equation}
for $k = 0, \ldots, N-1$. For example, if $N=5$ we have $\theta_0 = 0, \theta_1 = 0.25, \theta_2 = 0.50, \theta_3 = 0.75$ and $\theta_4 = 1.00$. In our experiments, we have fixed $N=20$. To average the results, we define
\begin{equation}
1-\textrm{MAE} = \frac{2}{(N+1)N} \sum_{k=0}^N \left( N - k \right) \left(1-\textrm{MAE}^{\theta_k}\right),
\end{equation}
and
\begin{equation}
\textrm{coverage} = \frac{2}{(N+1) N} \sum_{k=0}^N \left( N - k \right) \textrm{coverage}^{\theta_{k}}.
\end{equation}

\subsection{Experimental Results}

Once the experimental environment is defined, we proceed to compare the \ac{BeMF} recommender using the recklessness regularization (blue graphs in the subsequent figures) and without using it (orange graphs in the subsequent figures). 

The initial experiment aimed to find the optimal hyperparameter set for each recommender. To achieve this objective, we utilized genetic algorithms to tune the hyper-parameters within our model. This approach enabled us to efficiently navigate the extensive hyper-parameter space while mitigating the computational complexities linked with methods such as grid search. Moreover, genetic algorithms offer notable advantages in tackling multi-objective optimization challenges, a characteristic particularly relevant in our scenario. These algorithms has been configured within the parameters included in \cref{tab:genetic-algorithms}. The fitness of each individual has been computed by running a 5-fold cross validation over the training set in order to compute the averaged $1-\textrm{MAE}$ and $\textrm{coverage}$ scores of each fold. The implementation of these algorithms was carried out using the Java Jenetics\footnote{\url{https://jenetics.io/}} library.  

\begin{table}[ht]
\centering
\scriptsize
\begin{tabular}{|c|c|}
    \hline
    \textbf{Parameter}                   & \textbf{Value}         \\ \hline
    Population                  & 100           \\ \hline
    Number of generations       & 150           \\ \hline
    Selection operator          & Tournament    \\ \hline
    Crossover operator          & Recombination \\ \hline
    Survivor selection operator & NSGA2         \\ \hline
    Mutation probability        & 0.01          \\ \hline
\end{tabular}
\caption{Main parameters of the genetic algorithms used for search the optimal set of hyper-parameters.}
\label{tab:genetic-algorithms}
\end{table}

\begin{figure}[ht]
    \centering
    \includegraphics[width=\textwidth]{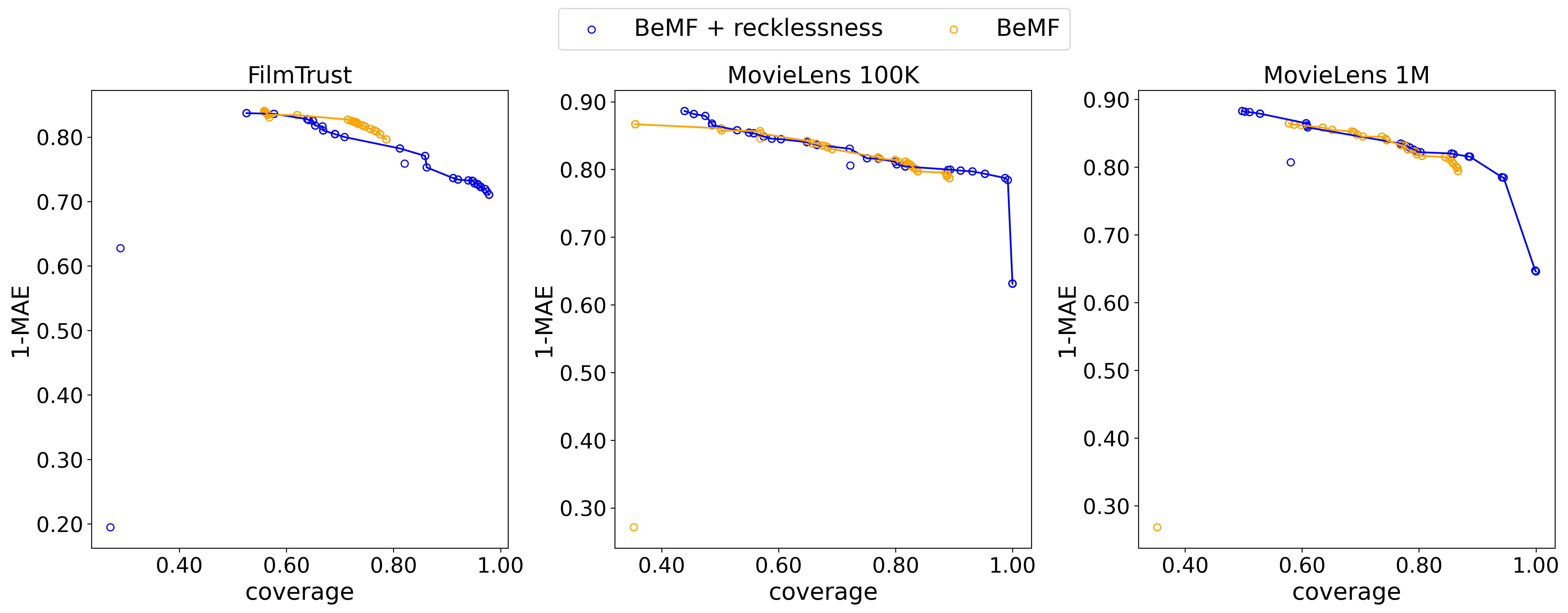}
    \caption{Pareto front obtained thorough hyper-parameters optimization using genetic algorithms.}
    \label{fig:pareto}
\end{figure}

\Cref{fig:pareto} shows the quality of the individuals from the last generation (graph points) as well as the Pareto front formed by these individuals (graph line). It can be observed that in all the three datasets analyzed, the Pareto front of the model using the recklessness regularization is consistently better since the Pareto front is wider. This means that the model employing recklessness regularization can achieve the same results as the model that does not use it and, moreover, provide more accurate predictions, sacrificing more coverage, or provide more coverage sacrificing the certainty of the predictions.

\begin{figure}[ht]
    \includegraphics[width=\textwidth]{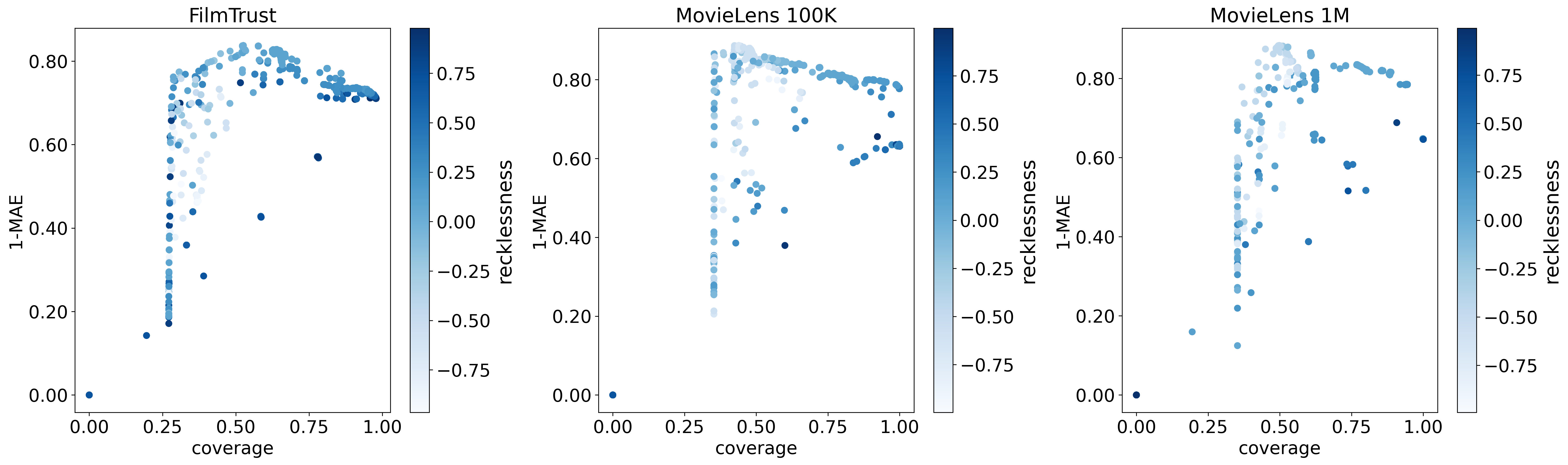}
    \caption{Recklessness value of each individual evaluated during the hyper-parameters optimization using genetic algorithms.}
    \label{fig:recklessness-value}
\end{figure}

This experiment demonstrates the advantage of using the recklessness regularization. However, it does not show the influence of this regularization on the balance between prediction quality and quantity. \Cref{fig:recklessness-value} displays the value of the recklessness regularization hyper-parameter for all the individuals evaluated by the genetic algorithms. It can be observed that, in general, positive values of recklessness provide a model with many predictions but less accuracy, whereas negative values of recklessness offer a model with fewer predictions but much higher accuracy.

This is consistent with the design of the recklessness regularization (see \cref{sec:recklessness}), as negative values imply that the predicted probability distributions by the model have a smaller variance, are flatter and thus are closer to a uniform distribution. This makes the model more conservative and only provides high reliabilities when it is very certain about the prediction issued. On the contrary, positive values achieve the opposite effect, causing the model to predict spiky probability distributions with high variance and forcing it to issue riskier predictions.

\begin{figure}[ht]
    \includegraphics[width=\textwidth]{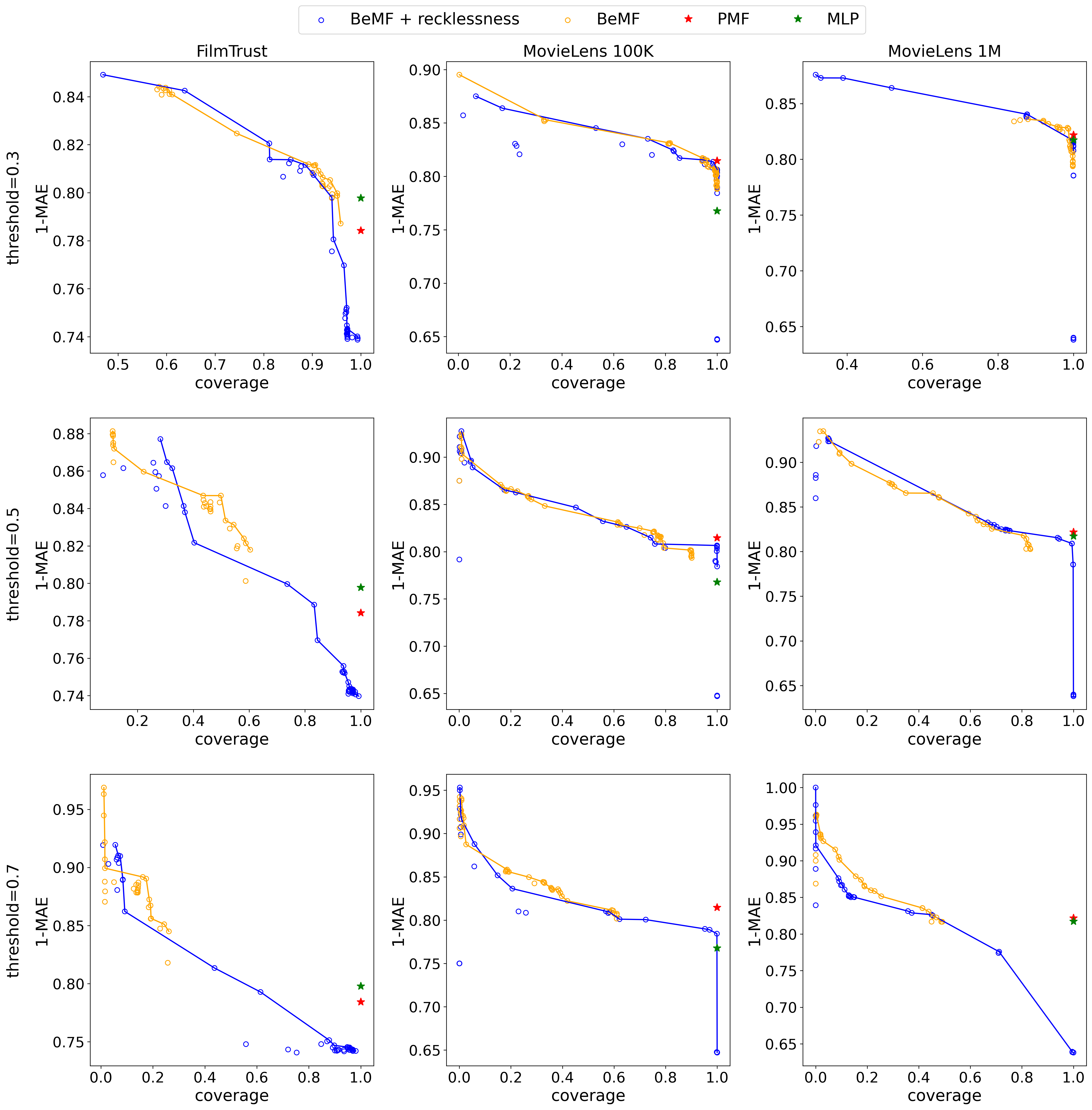}
    \caption{Test error of the solutions of the Pareto front found by the genetic algorithm.}
    \label{fig:test-error}
\end{figure}

These results evidence an improvement in the \ac{BeMF} model when the recklessness regularization is added. However, all these results have been calculated using validation sets, so overfitting may have occurred in spite of using cross-validation. \Cref{fig:test-error} displays the test results for the Pareto front individuals obtained with the genetic algorithm hyper-parameters optimization. Each dataset is presented in a column of the figure and each row shows the outcome of the quality measures, filtering predictions with a reliability below the threshold ($\theta$) indicated on the graph. As in \cref{fig:pareto}, the points represent the results of each individual in the Pareto front, while the lines depict the new Pareto fronts obtained in the test set. Additionally, two new reference models highly popular in the field of \ac{CF} have been added to this figure: PMF \cite{mnih2007probabilistic} and MLP \cite{he2017neural}. It is worth noting that these models do not provide a probability distribution as output since they act as regressors, resulting in predictions always being continuous values and the coverage always being of the 100\%. Therefore, their hyperparameters have been optimized through grid search for this reason instead of using a genetic algorithm.

The overall trend of the error in testing mirrors what was observed with validation error: when reckless regularization is applied, the Pareto front widens, allowing for a better alignment with the desired output type from the \ac{RS}. In general, the model with reckless regularization consistently achieves the most accurate predictions and the highest quantity of predictions. Furthermore, this trend is reinforced as less reliable predictions are discarded. On the other hand, concerning the baselines, the proposed model delivers superior predictions at the expense of coverage. Put differently, if the quality of baseline predictions falls short, the proposed model can enhance it provided one is willing to sacrifice less reliable predictions.

\begin{figure}[ht]
    \includegraphics[width=\textwidth]{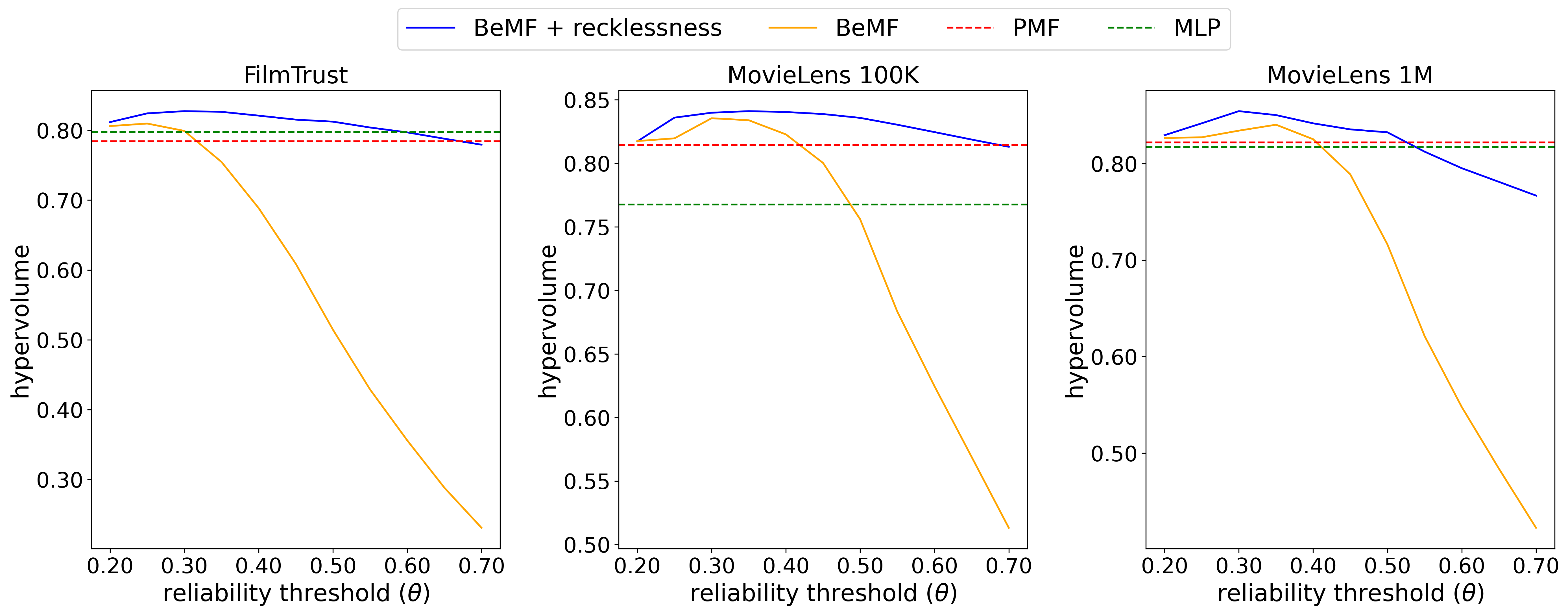}
    \caption{Variation of the Pareto front's hyper-volume in test when filtering all predictions with a reliability lower than the threshold specified.}
    \label{fig:hv}
\end{figure}

As a final analysis, we have compared these Pareto fronts by using their hyper-volume measure, i.e.\ the volume of the dominated portion of the objective space. This comparison is shown in \cref{fig:hv}, where the hyper-volume has been calculated both for model with and without recklessness when we filter out predictions with a reliability below the threshold indicated on the horizontal axis of each figure. To this figure, the hyper-volume of the PMF and MLP baselines has also been added. It can be observed that the model with reckless regularization consistently achieves the best results in most cases. Compared to the model without recklessness, the proposed model consistently outperforms due to the wider Pareto front, which allows it to cover more solutions and thus increase the hyper-volume. Regarding the baselines, the proposed model tends to offer superior results. In general, for low threshold values, the proposed model always performs better. Occasionally, when the threshold is very high and, therefore, the coverage of the proposed model is low, its hyper-volume may be lower than that of the baselines.

\section{Conclusions and Future Work} \label{sec:conclusions}

In this article, we have introduced a new regularization term for the cost functions of probability-based \ac{RS}. This new regularization allows us to adjust the variance of the underlying model's distribution. Negative values of this hyper-parameter results in models with lower variance and providing fewer but highly reliable predictions, while positive values lead to models with higher variance and offering more predictions, though of lower quality.Furthermore, this novel hyperparameter enhances the performance of the \ac{RS} in terms of both \textrm{MAE} and \textrm{coverage}, surpassing the performance of the same model lacking the hyperparameter as well as state-of-the-art models. All the described results have been validated using three different datasets, and the source code has been provided to ensure reproducibility.

Although the hyper-parameter has only been tested on one \ac{RS} model, its application in other systems appears promising. Therefore, future work includes evaluating the inclusion of recklessness in other \ac{CF} systems based on \ac{MF}. Similarly, applying the concepts of recklessness to \ac{CF} models based on neural networks~\cite{bobadilla2022neural,he2017neural} or neighborhood-based~\cite{singh2020movie} approaches is also considered.

\section*{Acknowledgments}

The first, second and forth named authors have been partially supported by \textit{Ministerio de Ciencia e Innovación} of Spain under the project PID2019-106493RB-I00 (DL-CEMG) and the \textit{Comunidad de Madrid} under \textit{Convenio Plurianual} with the Universidad Politécnica de Madrid in the actuation line of \textit{Programa de Excelencia para el Profesorado Universitario}. 

The third named author has been partially supported by the Madrid Government (\textit{Comunidad de Madrid – Spain}) under the Multiannual Agreement with the \textit{Universidad Complutense de Madrid} in the line Research Incentive for Young PhDs, in the context of the V PRICIT (Regional Programme of Research and Technological Innovation) through the project PR27/21-029, by the\textit{ Ministerio de Ciencia e Innovaci\'on} Project PID2021-124440NB-I00 (Spain) and by the BBVA Foundation grant COMPLEXFLUIDS.



 \bibliographystyle{elsarticle-num} 
 \bibliography{cas-refs}





\end{document}